\documentclass[aps,%
reprint,
superscriptaddress,
amsmath,amssymb,floatfix,pra,
noeprint
]{revtex4-2}
\usepackage{graphicx}
\usepackage{dcolumn}
\usepackage{comment}
\usepackage{bm}
\usepackage{xcolor}
\definecolor{darkblue}{rgb}{0,0,0.6}
\usepackage{hyperref}
\hypersetup{colorlinks,linkcolor=darkblue,citecolor=darkblue,urlcolor=darkblue}

\usepackage[normalem]{ulem}

\begin{document}

\title{Hyperuniformity in active fluids reshapes nucleation and capillary-wave dynamics}

\author{Raphaël Maire}
\email{maire@ub.edu}
\affiliation{Department of Condensed Matter, University of Barcelona, 08028 Barcelona, Spain}

\date{\today}

\begin{abstract}
While nucleation in typical active and driven fluids often appears equilibrium-like, striking departures emerge when large-scale fluctuations are strongly suppressed. Here, we investigate nucleation in nonequilibrium hyperuniform fluids by projecting the full density-field dynamics onto relevant collective variables. We demonstrate that nucleation is governed by a nonequilibrium quasi-potential rather than the reversible work of formation. Surprisingly, because of the reduced hyperuniform fluctuations, the nucleation probability no longer separates into the usual surface and volume contributions. Furthermore, accounting for capillary waves reveals a clear breakdown of detailed balance driven by nonreciprocal dynamics. More broadly, our framework can be readily extended to identify nonequilibrium signatures in conventional active fluids.
\end{abstract}

\maketitle

In equilibrium liquids, classical nucleation theory describes droplet formation in a supersaturated fluid through the reversible work $W(R)$ required to form a spherical nucleus of radius $R$ within it. The probability of observing such a nucleus is~\cite{onuki2002phase}:
\begin{equation}
    P(R)\propto e^{-W(R)/k_{\rm B}T}, \quad W(R)=\gamma A(R)-\Delta fV(R),
    \label{eq:work}
\end{equation}
where $T$ is the temperature, $\gamma$ the surface tension, $\Delta f$ the bulk free-energy density difference between phases, and $A(R)$ and $V(R)$ the nucleus area and volume. The first term is the interfacial cost, and the second is the bulk driving term favoring the new phase.

External forcing can strongly modify nucleation kinetics, but the same interfacial-bulk balance controlling the nucleation probability often remains visible, such as in fluids under shear flow~\cite{goswami2021homogeneous,mura2016effects}, ultrasound~\cite{nalesso2019review}, optical fields~\cite{alexander2019non}, or magnetic fields~\cite{wang2017strong}. Whether this apparent robustness survives in internally driven fluids, where nonequilibrium effects are intrinsic rather than externally imposed, is a much stricter test that active matter naturally provides.

In scalar active systems, for example, clear nonequilibrium effects do emerge, including negative effective surface tension~\cite{cates2024active} and unconventional pathways connecting nuclei to homogeneous or phase-separated states~\cite{Zakine2023,yao2025interfacial}. Yet the overall classical nucleation structure appears surprisingly robust~\cite{richard2016,redner2016,langford2024mechanics,Levis2017,cho2023nonequilibrium}: the nucleation statistics can often be described by an effective nucleation potential containing interfacial and bulk contributions~\cite{cates2023}, and in some cases additional active terms~\cite{ziethen2023nucleation}. At a geometric level, the appearance of volume and area terms is almost tautological: bulk effects are extensive in the nucleus volume, while interfacial effects are extensive in its surface area. Geometry alone, however, does not determine the nucleation probability, which also depends on the fluctuation statistics that produce the nucleus. When fluctuations are long-range correlated, large coherent variations over extended regions can be strongly suppressed, so forming a nucleus of radius $R$ requires an increasingly coordinated fluctuation. The associated statistical penalty, therefore, need not follow geometry: even if the deterministic interfacial cost scales with area, the rarity of the required fluctuation can grow faster than this area. The central question is thus whether geometric bulk and interfacial terms still control the nucleation probability once nonequilibrium fluctuation statistics are included. Active systems that can be effectively described by colored noise provide an example of this kind~\cite{maggi2022critical}. In such systems, nonequilibrium driving is expected to qualitatively alter the nucleation dynamics, but only up to a crossover scale. Beyond that scale, the dynamics become effectively governed by white noise, and the usual bulk and interfacial contributions should be recovered. To obtain a sharper breakdown of this picture, one must instead consider systems with much stronger suppression of large-scale fluctuations: hyperuniform systems.

Hyperuniform systems exhibit long-range correlations that suppress bulk density fluctuations, so the variance in particle number grows more slowly than the volume of the observation region~\cite{torquato2016hyperuniformity}. They thus realize precisely the scenario outlined above: nucleating a large droplet requires a coherent fluctuation over an extended region, yet such fluctuations are anomalously rare. As a result, the geometric volume and area scalings of bulk and interfacial contributions need not carry over directly to the scaling of the nucleation probability. The same suppression of large-scale fluctuations also underlies several striking nonequilibrium phenomena in hyperuniform fluids, including violations of the Mermin–Wagner theorem~\cite{PhysRevLett.131.047101,enhancing2024Maire,kuroda2024long,ikeda2024harmonic,keta2024long}, suppression of capillary waves~\cite{Maire2025b}, and a reduced upper critical dimension in $O(N)$ models~\cite{gao2025liquidgascriticalityhyperuniformfluids,ikeda2023correlated}.

A further subtlety is that a breakdown of the classical nucleation scalings does not, by itself, imply a breakdown of detailed balance in the reduced dynamics. Indeed, as noted in Ref.~\cite{cates2023}, when the nucleus is described only by its radius, the stochastic evolution can still be recast as an effective one-dimensional gradient dynamics, and thus exhibits an equilibrium-like structure. To expose genuinely nonequilibrium signatures, we must go beyond this minimal description and include additional slow degrees of freedom such as capillary waves. As we show below, this extended description brings out a nonreciprocal coupling between radial growth and interfacial fluctuations, which breaks detailed balance. The same logic applies more broadly to typical active systems, for example, those described by Active Model B+.

This work is divided as follows. We introduce in Sec.~\ref{sec:HU} the class of nonequilibrium hyperuniform fluids considered here and their coarse-grained field theory. We then derive in Sec.~\ref{sec:projection}, stochastic equations for the droplet radius and center of mass motion, by projecting the density field onto these collective coordinates. Next, in Sec.~\ref{sec:nucleation} we analyze single-droplet nucleation and show how hyperuniform fluctuations modify the effective nucleation quasi-potential, as well as the diffusion of droplets. We then extend the theory to include capillary-wave modes in Sec.~\ref{sec:CW}, which allows us to identify a genuine breakdown of detailed balance in the coupled coarse-grained dynamics. Finally, we quantify this irreversibility through entropy production and pathwise considerations in Sec.~\ref{sec:entropy}, before concluding and discussing further avenues in Sec.~\ref{sec:discussion}.

\section{Hyperuniform systems}
\label{sec:HU}

Hyperuniform fluids are unusual because they display long-range correlations, which in equilibrium typically require long-range interactions. Nucleation in such fluids is therefore an interesting and, to our knowledge, still largely unexplored problem~\footnote{Long-range forces can be naturally incorporated into the framework we developed by modeling them through a fractional Laplacian in the Landau free energy~\cite{kardar2007statistical}.}. Our focus, however, is on a different and more striking situation: systems with only short-range interactions whose nonequilibrium dynamics nevertheless generate long-range suppressed correlations. This behavior arises in many settings. See Refs.~\onlinecite{Maire2025} and \onlinecite{lei2024non}, and references therein for an extensive list.

In this work, we focus on a subset of such systems that become hyperuniform through nonequilibrium dynamics, specifically those in which the center of mass motion is fixed or damped and motion arises from momentum-conserving interactions~\cite{Maire2025}. This class includes, for example, random-organization models~\cite{hexner2015hyperuniformity, tjhung2015hyperuniform,anand2025emergent, galliano2026glass, yxch-hpbv}, active spinners~\cite{MassanaCid2021}, chiral particles~\cite{Lei2019,kuroda2023microscopic}, and particles with oscillating radius~\cite{ikeda2024harmonic}. Hyperuniformity in these systems requires two conditions: (i) there must be no external noise, because otherwise the center of mass is no longer bounded, diffuses, and hyperuniformity is destroyed~\cite{hexner2017noise}. (ii) the damping must remain linear, since nonlinear effects generate an effective external noise under the renormalization group~\cite{Maire2025}.

At large length scales, these systems are described by the hyperuniform Model B equation~\cite{Maire2025} for the density field $\rho(\bm r, t)$:
\begin{equation}
    \partial_t\rho =-\bm\nabla\cdot\left(-\Gamma(\rho)\bm\nabla\dfrac{\delta F}{\delta \rho}+\bm J_{\rm neq}\right) + \bm \nabla^2\left(\sqrt{2D(\rho)}\zeta\right),
\end{equation}
where $\Gamma$ and $D$ are the mobility and diffusivity respectively, $F$ is a Landau free energy functional, $\bm J_{\rm neq}$ collects the remaining possible non-variational terms~\cite{burekovic2026active} and $\zeta$ is a white Gaussian noise with unit variance. The key feature of this equation is the noise term that enters through a Laplacian, reflecting center of mass conservation in the overdamped dynamics (and it can be traced back, in an underdamped description, to a fluctuating stress). This is a strong departure from equilibrium fluctuating hydrodynamics, where the conserved noise enters as the divergence of a random current. Such an equation is quite general, as it describes active systems with short-range, reciprocal interactions and no added external noise, the latter of which would contribute an additional divergence noise term. Most hyperuniform fluids observed in nonequilibrium settings without fine-tuning can be described within this framework, where the noise arises from the interplay between damping (or overdamped dynamics) and motion induced by reciprocal interactions~\cite{Maire2025,lei2024non}. This excludes, for instance, systems with long-range interactions or at the critical point of the conserved directed percolation universality class. For such systems, we expect the conclusions of our analysis to remain qualitatively valid, though not quantitatively, since the underlying mechanisms leading to hyperuniformity are different.

That difference in noise structure has a direct consequence for density fluctuations. Writing the Fourier transform of the density field as $\rho(\bm k)$ the linear theory gives~\cite{hexner2017noise}: $\langle \delta\rho(\bm k)\delta\rho(-\bm k)\rangle \sim k^2$. Hence, the structure factor vanishes as $k\to 0$ and long-wavelength density fluctuations are strongly suppressed. By contrast, in equilibrium, one expects $\langle \delta\rho(\bm k)\delta\rho(-\bm k)\rangle\sim T$. This suggests a useful formal analogy~\cite{Maire2025, lei2019hydrodynamics}, where each mode is thermalized by an effective temperature $T_{\rm eff}(\bm k)\propto  k^2$, so that long-wavelength modes are effectively ‘‘colder'', and therefore less fluctuating than short-wavelength ones. This is not a thermodynamic temperature, but a compact way to express the suppression of large-scale fluctuations. Since a nucleus of size $R$ necessarily involves modes with $k\lesssim 1/R$, its formation is then suppressed not only by geometric bulk and interfacial costs, but also by the anomalously weak fluctuations available at those scales.

\section{Derivation of nucleation dynamics equation}
\label{sec:projection}

It remains to derive effective equations for the droplet degrees of freedom from the full density-field dynamics. To do so, we loosely follow Refs.~\onlinecite{lutsko2018systematically, cates2023, Lutsko2012, lutsko2013classical, blomker2020stochastic}.

\subsection{Setup}
\begin{figure*}
    \centering
    \includegraphics[width=0.8\linewidth]{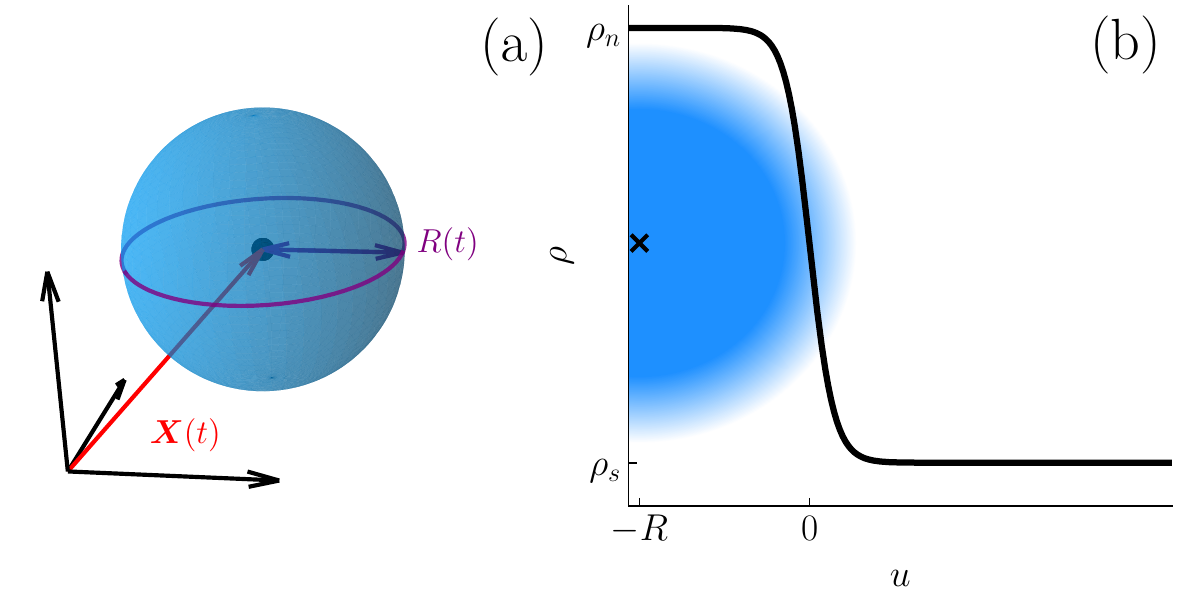}
    \caption{Cartoon summarizing the parametrization Eq.~\eqref{eq:param} (a) Position $\bm X(t)$ of the droplet with radius $R(t)$. (b) Parametrization of the density field along $u(\bm  r,t)=|\bm  r-\bm  X(t)|-R(t)$. }
    \label{fig:cartoon}
\end{figure*}

Although our theory can include density-dependent transport coefficients~\cite{lutsko2018systematically}, we take the mobility and diffusivity to be constant to isolate the effect of suppressed density fluctuations on nucleation and rescale space and time to absorb $\Gamma$. We also set the nonequilibrium current to zero because it mainly dresses the deterministic terms such as the effective surface tension and bulk contribution without changing the structure of the effective potential~\cite{cates2023}, contrary to the Laplacian noise. Finally, to compare with a conventional equilibrium-like fluid, we add an independent divergence noise. The resulting density dynamics in $d$ dimensions are:
\begin{equation}
    \partial_t\rho =\bm\nabla^2\left(\tilde\mu + \sqrt{2D}\zeta\right)+\bm\nabla\cdot\left(\sqrt{2D^{(b)}}\bm \xi\right),
        \label{eq:full_dynamics}
\end{equation}
where $\bm \xi$ is a unit variance white Gaussian noise, and $D^{(b)}$ is a diffusivity which can be thought to arise because the nonequilibrium system is put in contact with an external bath. This extra term restores ordinary density fluctuations on large scales, and equilibrium Model B would be recovered in the limit $D=0$ and $D^{(b)}\neq0$. We define the shifted variational chemical potential~\cite{chaikin1995principles}, $\tilde \mu(\bm r)=\frac{\delta F}{\delta \rho(\bm r)}-\mu_s$, where $F$ is a Landau free energy functional:
\begin{equation}
    F = \int \left(f(\rho(\bm r))+\dfrac{\kappa}{2}\left|\bm \nabla\rho\right|^2\right) d\bm r,
\end{equation}
with $\kappa$ a capillary coefficient penalizing density gradients, $f$ a non-convex local free-energy density that supports phase separation, and $\mu_s$ the chemical potential at infinity. We note that we shifted the chemical potential by the constant $\mu_s$. It does not affect the evolution equation, since it vanishes under the Laplacian, but it will be useful later in eliminating boundary terms from integration by parts~\footnote{It corresponds to a shift of the Landau free energy, $\tilde F = F - \mu_s \int \rho(\bm r) d\bm r$, where $\mu_s$ is the Lagrange multiplier enforcing mass conservation~\cite{chaikin1995principles}. In the evolution equation, it is often unnecessary to work explicitly with $\tilde F$, since the Laplacian preserves mass (and therefore removes this constant shift). However, introducing $\tilde F$ can be useful in some contexts. For example, the stationary profile is determined by $\delta \tilde F / \delta \rho = 0$, rather than by $\delta F / \delta \rho = 0$. As a corollary, the binodals are selected by the underlying minima of $\tilde F$, not of $F$.}.

We now assume the density field contains a single approximately spherical droplet of radius $R(t)$ centered at $\bm X(t)$, and parametrize it as
\begin{equation}
    \rho(\bm  r,t)=g(u(\bm  r,t)),\quad u(\bm  r,t)=|\bm  r-\bm  X(t)|-R(t).
    \label{eq:param}
\end{equation}
By construction, $u=0$ at the interface, $u<0$ inside the droplet, and $u>0$ outside. The function $g(u)$ is the radial interfacial profile which interpolates between the nucleus density $\rho_n$ and the outer supersaturated density $\rho_s$. In principle, its mean field shape is obtained from the stationary Euler–Lagrange equation $\delta \tilde F/\delta\rho=0$, with possible non-mean field corrections~\cite{kopf2008interfacial}. For the derivation below, it is enough to work in the thin-interface limit, where a sharp step approximates the profile,
\begin{equation}
g(u)= \rho_s+\Delta\rho\Theta(-u),\quad \Delta\rho=\rho_n-\rho_s,
\label{eq:sharp_interface}
\end{equation}
with $\Theta$ the Heaviside function, $\rho_n$ the nucleus density, and $\rho_s$ the density of the supersaturated fluid, far from the nucleus. Alternatively, we may ask that
\begin{equation}
\int b(r)g'(u)dr\simeq -\Delta\rho  b(R(t)),
\end{equation}
for any function $b$ and where the integral is over the radial coordinate only. Finally, in principle $\rho_n$, and more generally $g$, should depend on $R(t)$. We assume these contributions to be subdominant. In any case, one can always choose $R$ large enough for our theory to apply. We also note that the ansatz Eq.~\eqref{eq:param} \emph{cannot} preserve the local mass conservation imposed by the density field equation. This is not problematic in what follows, especially since we already accounted in a rough way for mass conservation via $\mu\to\mu-\mu_s$, but it should be kept in mind. The parametrization of the density field is summarized with two cartoons in Fig.~\ref{fig:cartoon}.

\subsection{Projection}

We now need to project the full dynamics of the density field onto the relevant slow variables, $\bm X(t)$ and $R(t)$, to obtain their time evolution.

We collect the slow variables into a vector:
\begin{equation}
    q_\alpha(t)\in\{R(t), X_1(t),\dots,X_d(t)\}.
\end{equation}
With this notation, the time evolution of $\rho$ naturally writes as
\begin{equation}
    \partial_t \rho = \sum_\beta \dot q_\beta \partial_\beta \rho=\sum_\beta\dot q_\beta\psi_\beta,
    \label{eq:linear_system}
\end{equation}
where we defined $\psi_\beta\equiv \partial_\beta \rho\equiv \partial_{q_\beta}\psi$ and easily find:
\begin{equation}
    \psi_R=\partial_R\rho=-g'(u),\quad\psi_{X_i}=\partial_{X_i}\rho=-g'(u)n_i,
    \label{eq:definition_psi}
\end{equation}
with $\bm  n(\bm  r,t)={(\bm  r-\bm  X(t))}/{|\bm  r-\bm  X(t)|}$.

To obtain closed equations for the collective coordinates $q_\beta$, we must choose a set of functions and project the field dynamics onto them~\footnote{Any invertible choice of test functions defines a legitimate closure, but different choices generally lead to different reduced equations. For a conserved (Model B) dynamics, a direct projection onto $\psi_\alpha=\partial_{\alpha}\rho$ probes $\langle \psi_\alpha,\bm\nabla^2\tilde\mu\rangle=\langle \bm\nabla^2\psi_\alpha,\tilde\mu\rangle$, i.e., a local interfacial moment of $\tilde\mu$. This overlooks the nonlocal bulk diffusion, which is essential for capillary waves~\cite{bray2001interface}, for example. The choice made below, $\bm\nabla^2\chi_\alpha=\psi_\alpha$, is therefore not merely convenient as it effectively inverts the Laplacian before projection and provides the proper non-local physics~\cite{bray2001interface}.}.
For the conserved dynamics considered here, a natural and convenient choice is to define $\chi_\beta$ by
\begin{equation}
    \bm\nabla^2 \chi_\beta = \psi_\beta,
\end{equation}
since under the scalar product:
\begin{equation}
    \langle a,b\rangle \equiv \int a(\bm  r)b(\bm  r) d\bm r,
\end{equation}
we have
\begin{equation}
    \begin{split}
    \langle \chi_\beta,\bm\nabla^2 f\rangle &= \langle \bm\nabla^2\chi_\beta,f\rangle = \langle \psi_\beta,f\rangle= \langle \partial_\beta \rho,f\rangle,\\
    \quad \langle\chi_\beta,\bm\nabla\cdot \bm  f\rangle &= -\langle \bm\nabla\chi_\beta,\bm  f\rangle,
    \end{split}
\end{equation}
for any fields $\bm f$ and $f$ by integration by parts, and with the definition $\langle \bm a, \bm b\rangle\equiv \sum_i\langle a_i, b_i\rangle$. In turn, this means that after projecting the dynamics of $\rho$ onto $\chi$, we can eliminate the Laplacian straightforwardly, since, for example:
\begin{equation}
    \langle \chi_\beta,\bm \nabla^2(\tilde\mu + \sqrt{2D}\zeta)\rangle=\langle \psi_\beta,\tilde\mu + \sqrt{2D}\zeta\rangle.
\end{equation}

Turning to the full dynamics in Eq.~\eqref{eq:full_dynamics}, we project onto $\chi_\alpha$, and together with Eq.~\eqref{eq:linear_system} we obtain:
\begin{equation}
    \begin{split}
    \sum_\beta\Big\langle \chi_\alpha,\psi_\beta\Big\rangle\dot q_\beta=&\langle \psi_\alpha,\tilde\mu[\rho]\rangle
    +\sqrt{2D}\langle \psi_\alpha,\zeta\rangle\\
    &-\sqrt{2D^{(b)}}\langle \bm\nabla\chi_\alpha,\bm\xi\rangle.
    \end{split}
    \label{eq:projection}
\end{equation}
We define the symmetric matrix:
\begin{equation}
    G_{\alpha\beta}\equiv\langle \chi_\alpha, \psi_\beta\rangle,
\end{equation}
since $\bm X$ is a vector and $R$ is a scalar, we immediately obtain $G_{RX_i}=G_{X_iR}=0$. Moreover, by isotropy of space, we get $G_{X_iX_j}\equiv\delta_{ij}G_{XX}$. 

We finally write the evolution equations for our slow fields by inverting $G$:
\begin{equation}
     \dot q_\beta=\dfrac{\langle \psi_\beta,\tilde\mu[\rho]\rangle
    +\sqrt{2D}\langle \psi_\beta,\zeta\rangle
    -\sqrt{2D^{(b)}}\langle \bm\nabla\chi_\beta,\bm\xi\rangle}{G_{\beta\beta}},
    \label{eq:dynamics_slow_field_before_integral}
\end{equation}
where we used the fact that $G$ is a diagonal matrix in our simple setting.

\subsection{Resulting equations}

We evaluate all integrals entering the scalar product (Eq.~\eqref{eq:dynamics_slow_field_before_integral}) in an arbitrary dimension $d$ in Appendix~\ref{sec:integrals_first}. In the main text, we focus on the case $d=3$, while the case $d=2$ and its subtleties are deferred to Appendix~\ref{sec:d=2}.

The equation for the radius $\dot q_R\equiv\dot R$ is found to be:
\begin{equation}
\dot R=\frac{2\gamma_0}{\Delta\rho^2 R}\left(\frac{1}{R_c}-\frac{1}{R}\right)
+\sqrt{2\big(D_R(R)+D_R^{(b)}(R)\big)}\eta_R(t),
\label{eq:R_dot_main}
\end{equation}
where $\gamma_0$ is the mean field surface tension~\cite{onuki2002phase}, $R_c$ is the critical radius
\begin{equation}
R_c=\frac{2\gamma_0}{(\rho_n-\rho_s)f'(\rho_s)-\big(f(\rho_n)-f(\rho_s)\big)},
\label{eq:R_c_main}
\end{equation}
beyond which supercritical droplets grow deterministically, $\eta_R$ is a Gaussian white noise with unit variance and zero mean, and
\begin{equation}
D_R(R)= D \frac{\gamma_0}{4\pi \kappa} \frac{1}{\Delta\rho^4} \frac{1}{R^{4}},
\quad
D_R^{(b)}(R)= D^{(b)} \frac{1}{4\pi} \frac{1}{\Delta\rho^2} \frac{1}{R^{3}},
\label{eq:Diffusivities_R_main}
\end{equation}
are the noise strengths for the radius evolution, related to the Laplacian noise and divergence noise, respectively, in the density-field equation.

The center of the nucleus evolves as:
\begin{equation}
\dot{\bm X}
=\sqrt{2\big(D_X(R)+D_X^{(b)}(R)\big)} \bm\eta_X(t),
\label{eq:X_dot_main}
\end{equation}
where $\bm \eta_X$ is a Gaussian vector noise with unit variance and
\begin{equation}
D_X(R)= D \frac{\gamma_0}{4\pi \kappa} \frac{27}{\Delta\rho^4} \frac{1}{R^{4}},
\quad
D_X^{(b)}(R)= D^{(b)} \frac{9}{4\pi} \frac{1}{\Delta\rho^2} \frac{1}{R^{3}}.
\label{eq:diffusivities_X_main}
\end{equation}
are the translational noise strengths. All equations are to be interpreted in the Stratonovich sense~\cite{Lutsko2012}.

Both noise strengths arising from the Laplacian noise scale as $D_X,D_R\sim R^{-4}$, whereas the bath-induced terms decay more slowly $D_X^{(b)},D_R^{(b)}\sim R^{-3}$. The hyperuniform noise is therefore weaker than the equilibrium one at large $R$. Because the explicit drift term in the projected Stratonovich equation is unchanged, we must conclude that large-$R$ nuclei fluctuations are more strongly suppressed in a hyperuniform system than in an equilibrium-like one. We now show this explicitly.

\section{Nucleation of a hyperuniform system}
\label{sec:nucleation}
\subsection{Single droplet nucleation}

We first analyze the dynamics of $\dot R$, when $D_R=0$ and $D_R^{(b)}\neq0$, that is, the nucleation dynamics of an equilibrium system described by model B. In this case, we can rewrite Eq.~\eqref{eq:R_dot_main} as:
\begin{equation}
    \dot R=-\mathcal M_R^{(b)}(R)\dfrac{\partial W}{\partial R}+\sqrt{2D^{(b)}\mathcal M_R^{(b)}(R)}\eta_R(t),
\end{equation}
with $W(R)=4\pi R^2\gamma_0-(4/3)\pi R^3\Delta f$ the reversible work necessary to form a nucleus with $\Delta f\equiv(\rho_n-\rho_s)f'(\rho_s)-\big(f(\rho_n)-f(\rho_s)\big)$~\footnote{ In $\Delta f=(\rho_n-\rho_s)f'(\rho_s)-\big(f(\rho_n)-f(\rho_s)\big)$, the first term is the chemical work (the $\mu_s \Delta\rho$ contribution, with $\mu_s=f'(\rho_s)$) associated with exchanging mass with the surrounding metastable phase. Equivalently, this is the grand-potential free-energy change $\Delta g$, with $g=f-\mu_s\rho$ where $\mu_s$ is a Lagrange multiplier for mass conservation.}, and $\mathcal M_R^{(b)}(R)=D_R^{(b)}/D^{(b)}$. We immediately obtain the quasi-stationary nucleus-radius distribution
\begin{equation}
P(R)\propto e^{-W(R)/D^{(b)}},
\end{equation}
which involves, up to subexponential corrections, the reversible work, as expected at equilibrium.

We now turn to the case of interest, $D_R^{(b)} = 0$ and $D_R \neq 0$, corresponding to the hyperuniform system. The deterministic part of the droplet dynamics is unchanged, so the reversible work required to form a nucleus is the same as before. What changes is the fluctuation term, and therefore the statistical weight of nuclei of different sizes.

To make this explicit, it is convenient to rewrite Eq.~\eqref{eq:R_dot_main} in the form of an effective stochastic gradient flow,
\begin{equation}
    \dot R=-\mathcal M_R(R)\dfrac{\partial \tilde W}{\partial R}+\sqrt{2D\mathcal M_R(R)}\eta_R(t),
\end{equation}
where $\mathcal M_R(R)=D_R/D$ and:
\begin{equation}
\tilde W(R)=\frac{8\pi}{3}\kappa \Delta\rho^2 R^3-\pi\kappa \Delta\rho^2 \frac{\Delta f}{\gamma_0} R^4,
\label{eq:W_tilde}
\end{equation}
where $\tilde W$ reaches a maximum at $R_c$, and once again, the leading probability distribution is given by:
\begin{equation}
P(R)\propto e^{-\tilde W(R)/D}.
\end{equation}
The key point is that, as for active models~\cite{cates2023}, $\tilde W(R)$ is \emph{not} the work necessary to form the nucleus, but instead the quasi-potential~\cite{Touchette2009} that properly accounts for fluctuations once the non-free energy terms or radius-dependent noise amplitude are taken into account. As a result, the probability distribution is no longer organized by the usual geometric surface/volume scalings ($\sim R^2$ and $\sim R^3$), but instead by terms scaling as $R^3$ and $R^4$ (Eq.~\eqref{eq:W_tilde}). For example, the $R^3$ term still depends on $\kappa$, and thus retains an interfacial origin, but it cannot be interpreted as a standard surface-tension contribution. In contrast, $R^4$ depends on $\Delta f$, but also on an interfacial contribution given by $\kappa/\gamma_0$.

Physically, hyperuniformity suppresses large-scale density fluctuations, so the coherent fluctuations needed to build a large droplet become increasingly rare. In other words, even though the deterministic work of formation is unchanged, the fluctuations available to overcome that work weaken with increasing $R$. In a way, the fluctuations effectively \emph{dress} the nucleation work.

\subsection{Diffusion of droplets}

From Eq.~\eqref{eq:diffusivities_X_main}, the center of mass noise strength of a droplet is strongly reduced in the hyperuniform system compared with the regular one, since $D_X^{(b)}/D_X \sim R$. Large droplets in a hyperuniform setting are therefore less mobile than in a thermal one, and coarsening in a multi-nucleus system should be dominated by diffusive mass transfer rather than by droplet coalescence~\cite{onuki2002phase}. This naturally points to Ostwald ripening. In Lifshitz-Slyozov-Wagner theory~\cite{bray1994theory}, droplets with $R>R_c(t)$ grow while those with $R<R_c(t)$ shrink, and the dissolved material from smaller droplets feeds the larger ones. As coarsening proceeds, the global supersaturation decreases self-consistently, which causes the critical radius $R_c(t)$ to increase in time. To leading order, this mechanism is gradient-driven and only weakly affected by noise~\cite{bray1994theory}. Since the deterministic part of our dynamics is the same as in equilibrium, we expect the coarsening to follow the usual equilibrium behavior at first approximation, in particular $l(t)\sim t^{1/3}$, where $l$ is the typical droplet size.

\section{Nonequilibrium dynamics and capillary waves}
\label{sec:CW}
We have shown that the probability of a pre-critical nucleus is no longer governed by the usual surface–volume form, nor by the reversible work of formation. Yet the effective dynamics of $R$ remain equilibrium-like. Indeed, because the reduced description involves essentially a single slow variable (the droplet radius), its stochastic evolution can always be recast as a gradient flow with an effective detailed balance. To reveal genuinely nonequilibrium dynamical signatures, one must go beyond this one-variable description and include additional slow modes, such as capillary-wave fluctuations.

\subsection{Capillary waves}

\begin{figure}
    \centering
    \includegraphics[width=0.9\linewidth]{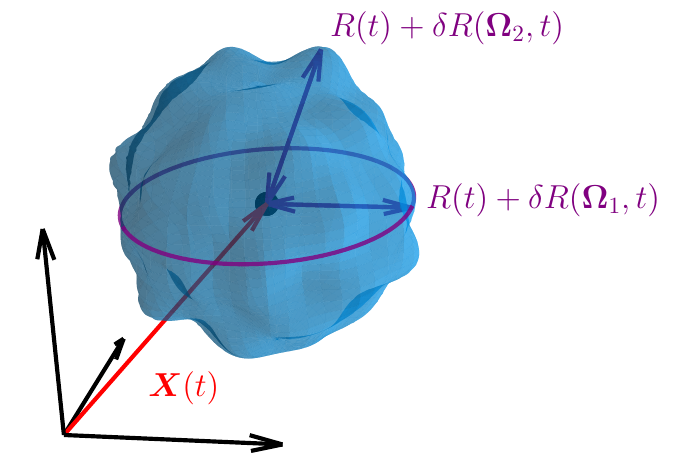}
    \caption{Cartoon summarizing the new parametrization Eq.~\eqref{eq:new_param}.}
    \label{fig:cartoon2}
\end{figure}

To include capillary waves, it is necessary to modify the ansatz by~\cite{besse2023interface, bray2001interface}:
\begin{equation}
    \begin{split}
    \rho(\bm  r,t)=&g(u(\bm  r,t)),\\
     u(\bm  r,t)=&|\bm  r-\bm  X(t)|-(R(t) + \delta R(\bm\Omega, t)),
    \end{split}
    \label{eq:new_param}
\end{equation}
where $\delta R(\bm \Omega, t)$ describes angular deformations of the interface and thus allows for capillary-wave modes. This new parametrization is described in Fig.~\ref{fig:cartoon2}.

We expand these deformations in hyperspherical harmonics:
\begin{equation}
    \delta R(\bm\Omega,t)=\sum_{\ell\ge 2}\sum_{J=1}^{N_{\ell,d}} a_{\ell J}(t)Y_{\ell J}(\bm\Omega),
\end{equation}
where $J$ is a collective index labeling the $N_{\ell, d}$ hyperspherical harmonics of degree $\ell$. For capillary waves in an isotropic space, the linear dynamics must depend only on $\ell$, while the $J$-degeneracy reflects different orientations of the same mode.

Within our projection framework, incorporating capillary waves simply amounts to enlarging the set of slow variables to include the amplitudes $a_{\ell J}$:
\begin{equation}
     q_\alpha(t)\in\{R(t), X_1(t),\dots,X_d(t), \{a_{\ell J}\}\},
\end{equation}
while Eq.~\eqref{eq:projection} still gives the reduced dynamics for $\dot q_\beta$.

We evaluate the required integrals in arbitrary dimension in Appendix~\ref{app:integrals_capillary}. This yields evolution equations for the capillary amplitudes coupled to the previously derived equations for $\bm X$ and $R$. In $d\neq 3$, there exists a nontrivial coupling between $R$ and $a_{\ell J}$. Since this does not affect the main point we will make, we focus on the simpler $d=3$ case, in which $\dot R$ remains unchanged and is still described by the previous equation, which does not incorporate capillary waves. We refer the reader to Appendix~\ref{sec:det_capillary} for details.

We therefore obtain in the $d=3$ linear regime:
\begin{equation}
\dot a_{\ell J}=-\Gamma_\ell(R)a_{\ell J}+\sqrt{2D_{\ell}(R)+2D^{(b)}_{\ell}(R)}\eta_{a_{\ell J}}(t),
\label{eq:capillary_mode}
\end{equation}
with
\begin{equation}
\Gamma_\ell(R)=\frac{\gamma_0}{\Delta\rho^{2}R^{3}} (2\ell+1) (\ell-1)(\ell+2),
\end{equation}
and the effective noise strengths
\begin{equation}
\begin{split}
D_{\ell}(R)&=D \frac{\gamma_0}{\kappa} \frac{(2\ell+1)^2}{\Delta\rho^{4}}\frac{1}{R^{4}},\\
D_{\ell}^{(b)}(R)&=D^{(b)} \frac{2\ell+1}{\Delta\rho^{2}} \frac{1}{R^{3}}.
\end{split}
\label{eq:D_ell_explicit}
\end{equation}

As a consistency check, we consider again the equilibrium limit $D=0$ with  $D^{(b)}\neq 0$. The dynamics can be written as a stochastic gradient flow:
\begin{equation}
    \dot a_{\ell J}
    =-\mathcal M_\ell^{(b)}(R)\frac{\partial U_{\rm cw}}{\partial a_{\ell J}}
    +\sqrt{2D^{(b)}\mathcal M_\ell^{(b)}(R)}\eta_{a_{\ell J}}(t),
\end{equation}
with $\mathcal M_\ell^{(b)}=D_\ell^{(b)}(R)/D^{(b)}$
\begin{equation}
U_{\rm cw}(\{a_{\ell J}\})
=\sum_{\ell\ge 2}\sum_J
\frac{\gamma_0}{2}(\ell-1)(\ell+2)\,a_{\ell J}^2.
\end{equation}
This is the standard quadratic capillary-wave Hamiltonian, i.e., the interfacial energy cost of small shape fluctuations~\cite{komura1993dynamical}. 

Since the system is equilibrium-like, we can find a \emph{global} potential $U$ such that~\cite{Fruchart2026}
\begin{equation}
    \dot q_{\beta}=-\mathcal M_\beta^{(b)} \dfrac{\partial U}{\partial q_\beta}+\sqrt{2D^{(b)}\mathcal M_\beta^{(b)}}\eta_\beta(t).
    \label{eq:coupled_eq}
\end{equation}
In the present case, because the radial mode and capillary modes decouple in $d=3$ (unlike in $d\neq 3$ where $U_{\rm cw}$ and $W$ depend on $R$ and $a_{\ell J}$, respectively), that potential is simply additive, 
\begin{equation}
    U(R,\{a_{\ell J}\}) = W(R)+U_{\rm cw}(\{a_{\ell J}\}),
\end{equation}

We now turn to the intrinsically nonequilibrium hyperuniform limit, where the density field is driven only by Laplacian noise ($D^{(b)}=0$ and $D\neq 0$). In this limit, the capillary dynamics \eqref{eq:capillary_mode} can still be written in gradient form
\begin{equation}
\dot a_{\ell J}
=-\mathcal M_\ell(R)\frac{\partial \tilde U_{\rm cw}}{\partial a_{\ell J}}
+\sqrt{2D\mathcal M_\ell(R)}\eta_{a_{\ell J}}(t),
\label{eq:hyperuniform_grad_a}
\end{equation}
with $\mathcal M_\ell=D_\ell(R)/D$ and
\begin{equation}
\begin{split}
\tilde U_{\rm cw}(R,\{a_{\ell J}\})
&=\sum_{\ell\ge 2}\sum_J \frac{\Gamma_\ell(R)}{2\mathcal M_\ell(R)}a_{\ell J}^2\\
&=\frac{\kappa\Delta\rho^2}{2}R\sum_{\ell\ge2}\sum_J
\frac{(\ell-1)(\ell+2)}{2\ell+1}a_{\ell J}^2.
\end{split}
\label{eq:U_CW_hyperuniform}
\end{equation}
This immediately reproduces the hyperuniform suppression of capillary fluctuations reported and measured in Ref.~\cite{Maire2025b} for flat interfaces:
\begin{equation}
\overline{a_{\ell J}^2}
=\frac{D_\ell(R)}{\Gamma_\ell(R)}
=\frac{D/R}{\kappa\Delta\rho^2}\frac{2\ell+1}{(\ell-1)(\ell+2)},
\label{eq:capillary_variance_hyperuniform}
\end{equation}
so that the variance scales as $1/\ell$ instead of the typical $1/\ell^2$ for thermal systems.

The key point we want to emphasize concerns the coupled behavior. Although the $R$ mode and each capillary mode can individually be written in gradient form
\begin{equation}
    \dot q_{\beta}=-\mathcal M_\beta \dfrac{\partial \tilde U_\beta}{\partial q_\beta}+\sqrt{2 D\mathcal M_\beta}\eta_\beta(t),
    \label{eq:coupled_neq}
\end{equation}
with $\tilde U_R=\tilde W$ and $\sum_{\ell,J}\tilde U_{a_{\ell J}}=\tilde U_{\rm cw}$, their joint evolution is genuinely nonequilibrium. In $d=3$, this is already clear from the asymmetric dependence of the potentials: $\tilde W(R)$ depends only on $R$, whereas $\tilde U_{\rm cw}(R, \{a_{\ell J}\})$ depends on both $R$ and $a_{\ell J}$. \emph{Under a stochastic evolution}, capillary waves are therefore driven by radial change, while $\dot R$ does not feel the capillary amplitudes. This one-way nonreciprocal coupling rules out a single scalar potential governing all coarse variables. In $d\neq 3$, where $\tilde W$ also depends on $\{a_{\ell J}\}$, the same conclusion follows from the more general failure of the integrability condition~\cite{Fruchart2026}:
\begin{equation}
\partial_{a_{\ell J}}\partial_R\tilde W \neq
\partial_R\partial_{a_{\ell J}}\tilde U_{\rm cw},
\label{eq:integrability}
\end{equation}
which shows that the forces cannot be obtained from a unique potential because the two variables have partially antagonistic goals.

\section{Quantifying Irreversibility}
\label{sec:entropy}

The coarse-grained dynamics are intrinsically nonequilibrium, so we now seek quantitative measures of this irreversibility. As a first indicator, we evaluate the entropy production in the metastable region.

\subsection{Entropy production}

Because the capillary-wave description is nonreciprocal, the coarse-grained dynamics breaks detailed balance. We therefore expect a finite (quasi-)stationary entropy production already in the metastable regime.

Importantly, even in the reduced description involving only the radius $R$, the entropy production can in principle remain nonzero when both noises are present with heat flowing from one to the other~\cite{enhancing2024Maire}. This contribution is not visible in our present formulation because we combined the two Gaussian noises into a single effective noise. Although this reduction is mathematically legitimate, it hides the distinction between the two ‘‘thermostats'' and therefore misses the entropy production associated with heat exchange between them whenever $D^{(b)}\neq D$~\cite{lee2018stochastic,murashita2016overdamped,van2010three}.

Regardless of this effect, we focus on the case purely driven by the hyperuniform noise ($D^{(b)}=0$), and define the Fokker-Planck equation~\cite{seifert2012stochastic} describing the coupled Langevin Eqs.~\eqref{eq:coupled_neq}:
\begin{equation}
\begin{gathered}
\partial_t P(\bm q,t)=-\sum_\beta \partial_\beta  \mathcal J_{\beta},\\
\mathcal J_\beta=\left[-\mathcal M_\beta \partial_\beta  \tilde U_\beta-\frac{D}{2}\partial_\beta  \mathcal M_\beta\right]P
-D\mathcal M_\beta\partial_\beta  P,
\end{gathered}
\end{equation}
with $P(\bm q)$ the joint probability distribution of all collective variables.

To make progress, we assume that the capillary modes relax quickly with respect to the radius dynamics. In this setting, the stochastic radius $R(t)$ acts as a protocol, dynamically fixed by $\dot R$ which is independent of the harmonic modes. This separation of timescales allows us to compute the capillary-wave entropy production as its excess contribution~\cite{hatano2001steady}.

Since the capillary modes obey an Ornstein-Uhlenbeck dynamics at fixed $R$, the stationary condition $\mathcal J_{a_{\ell J}}=0$ yields the conditional distribution:
\begin{equation}
P(\{a_{\ell J}\}|R)\propto \exp\left[-\frac{1}{2}\sum_{\ell, J} \frac{\kappa\Delta\rho^2}{D}\frac{(\ell-1)(\ell+2)}{2\ell+1}Ra_{\ell J}^2\right].
\label{eq:distribution_condi}
\end{equation}
We then factorize the joint distribution as:
\begin{equation}
P(\bm q)=P(R,\{a_{\ell J}\})=P_R(R)P(\{a_{\ell J}\}|R),
\end{equation}
where we discarded the diffusing $\bm X$ variable. Because $P(\{a_{\ell J}\}|R)$ depends on $R$, the radial current does not vanish even though $\mathcal J_{a_{\ell J}}=0$. We approximate $P_R(R)$ so that the $a$-independent part of $\mathcal J_R$ vanishes to focus only on the capillary waves. This yields:
\begin{equation}
\mathcal J_R=\frac{\mathcal M_R (R)}{2R}\left[\sum_{\ell, J}\left(\kappa\Delta\rho^2\frac{(\ell-1)(\ell+2)}{2\ell+1}R a_{\ell J}^2-D\right)\right]P.
\end{equation}

The steady-state entropy-production rate is~\cite{seifert2012stochastic}:
\begin{equation}
\dot S_{\rm tot}=\sum_\beta\int \frac{\mathcal J_\beta^2}{D\mathcal M_\beta P}d\bm q=\int \frac{\mathcal J_R^2}{D_R P}d\bm q,
\end{equation}
since $\mathcal J_{ a_{\ell J}}=0$. Averaging over the Gaussian capillary modes at fixed $R$ (Eq.~\eqref{eq:distribution_condi}) and using Eq.~\eqref{eq:Diffusivities_R_main}, gives
\begin{equation}
\begin{split}
\dot S_{\rm tot}&=\frac{N_{\rm cw}}{2}\int P_R(R)\frac{D_R(R)}{R^2}dR\\
&=\frac{N_{\rm cw}D\gamma_0}{8\pi\kappa\Delta\rho^4}\int P_R(R)R^{-6}dR\\
&\propto N_{\rm cw}\overline{R^{-6}}>0,
\end{split}
\end{equation}
with $N_{\rm cw}$ the total number of capillary-wave modes kept in the reduced description:
\begin{equation}
N_{\rm cw}\equiv \sum_{\ell=2}^{\ell_{\max}}\sum_{J=1}^{N_{\ell,d=3}}1
=\sum_{\ell=2}^{\ell_{\max}}N_{\ell,d=3}=\ell_{\max}^2+2\ell_{\max}-3.
\end{equation}

As expected, the entropy production is nonzero and depends not only on the droplet radius but also on the number of capillary-wave modes retained. Because each mode provides an additional channel for entropy production, the total entropy production diverges as $N_{\rm cw}\to\infty$. This behavior is standard in continuum descriptions~\cite{borthne2020time,markovich2021thermodynamics}. In practice, however, the theory is always understood to have an ultraviolet cutoff $\ell_{\rm UV}$ and $1/R_{\min}$, beyond which the coarse-grained description no longer applies.

\subsection{Pathwise irreversibility of nucleation}

The entropy production computed above probes irreversibility in the metastable quasi-stationary regime. Since our main interest is nucleation itself, it is useful to characterize irreversibility directly at the level of stochastic paths that eventually lead to nucleation.

The Langevin equations~\eqref{eq:coupled_neq} define stochastic paths $\bm q(t)$ in the space of collective variables. A ‘‘detailed nucleation'' event corresponds to a trajectory that starts at $t=0$ at a given point in the metastable basin and reaches the barrier $R=R_c$ at time $t=t_c$ with other variables fixed. In the weak-noise/high-barrier limit, the probability of such an event is exponentially dominated by a single trajectory, the most probable nucleation path, denoted $\bm q^{\rm nuc}_*(t)$. One may similarly define the reverse process as the relaxation from the barrier $R=R_c$ back into the previously defined initial point. In the same weak-noise limit, this relaxation follows the deterministic path, denoted $\bm q^{\rm rel}_*(t)$.

At equilibrium, detailed balance implies pathwise reversibility and the most probable nucleation path must be the time-reversal of the relaxation path~\cite{avanzini2024methods, Zakine2023},
\begin{equation}
\bm q^{\rm nuc}_*(t)=\bm q^{\rm rel}_*(t_c-t).
\end{equation}
Out of equilibrium, this equality will not generically hold. Notably, Eq.~\eqref{eq:integrability} is enough to rule out pathwise reversibility~\cite {Zakine2023}. A mismatch between $\bm q^{\rm nuc}_*(t)$ and $\bm q^{\rm rel}_*(t_c-t)$ therefore provides a transparent signature of irreversibility in the nucleation dynamics. 

Taking a step back, however, our nonequilibrium dynamics is a rather ill-suited example in the context of nucleation. Capillary waves are not expected to play a major role in the nucleation event itself, and although including them can produce irreversibility, this occurs only when one imposes artificial constraints on the initial and final states. In the physically relevant case $a_{\ell J}=0$ at both endpoints, the most probable path never leaves $a_{\ell J}(t)= 0$. More generally, the transition from any point in the metastable basin to $R=R_c$ splits into an activated trajectory for $R(t)$ and a purely relaxational evolution of the capillary modes $\dot a$, toward $a=0$, as shown in Appendix~\ref{app:relaxation}. Capillary waves, therefore, do not provide a meaningful nonequilibrium contribution to nucleation.

As shown in previous studies, the physically relevant nonequilibrium character of nucleation in active systems should instead be sought in variations of droplet morphology and density.~\cite{Zakine2024, yao2025non}. A natural way to capture this would be to include, for instance, $\rho_n$ as an additional dynamical variable~\cite{Duran-Olivencia_Yatsyshin_Kalliadasis_Lutsko_2018}, or to introduce a parameter that directly parametrizes $g$ itself.

\section{Discussion}
\label{sec:discussion}

Using projection methods, we derived evolution equations for a set of collective variables from the underlying density-field dynamics. This allowed us to show that, in hyperuniform fluids, nucleation is governed by a nonequilibrium quasi-potential rather than by the reversible work of formation. As a result, the nucleation exponent no longer follows the usual surface/volume decomposition. Including capillary modes further shows that the reduced dynamics are genuinely nonequilibrium, with nonreciprocal couplings and broken detailed balance.

A direct comparison with particle simulations is possible but not straightforward. Preliminary molecular-dynamics results are difficult to interpret quantitatively because the accessible nuclei are often small, and reaching the relevant regime would require advanced rare event sampling methods~\cite{Allen2009}. Our theory is expected to apply only at sufficiently large scales. First, it relies on a continuum description of the fluid, which should be reasonable only when the nucleus radius is at least about an order of magnitude larger than the mean free path. Second, it assumes that the density is the only relevant field, with other slow variables, such as angular momentum~\cite{gao2025liquidgascriticalityhyperuniformfluids}, having been adiabatically eliminated. The scales at which these additional effects become important must be assessed on a case-by-case basis. They may modify the density evolution equation through terms that cannot be derived from a free energy, as well as through temperature gradients or density-dependent activity. Some of these effects can be reduced by working in a parameter regime where the two phases have similar densities. This, however, introduces a different difficulty: critical fluctuations become important, and the thin-interface limit (where the nucleus size is assumed to be much larger than the interfacial width) breaks down because the surface tension decreases as the critical point is approached~\cite{onuki2002phase}. All of these effects compete with the asymptotic regime described by our theory. Although they could in principle be incorporated into our framework, doing so lies beyond the scope of the present work.

A natural next step is to apply the same projection method to Active Model B+, building on Ref.~\onlinecite{cates2023}. While capillary waves can be included as above, the more important extension is to identify and incorporate the relevant morphological degrees of freedom into the set of dynamical variables, to capture nonequilibrium signatures that are missed by a radius-only description and are likely more important than capillary modes for active nucleation pathways.

More broadly, the framework developed here is well-suited to extensions with long-range interactions and additional coupled fields such as polar and nematic order. It also allows for the study of other hyperuniform systems such as those generated by critical fluctuations~\cite{wiese2024hyperuniformity}. This framework provides a unified route into the largely unexplored physics of nonequilibrium nucleation in active matter.

\begin{acknowledgments}
I thank C. Nardini for discussions that motivated much of the work presented here. I also thank G. Foffi, F. Smallenburg, Y. Kuroda, L. Berthier and L. Galliano for valuable discussions on the physics of nucleation and hyperuniformity. Finally, I thank L. Sarfati for insightful comments on the mathematical formalism and problems underlying projection methods.
\end{acknowledgments}

\appendix

\section{Integrals}
\label{sec:integrals_first}
In this appendix, we compute the integrals arising in the projection of the density field onto the collective variables in Eq.~\eqref{eq:dynamics_slow_field_before_integral}. 
\subsection{\texorpdfstring{$\chi_\beta$}{chi}  and Laplacian inversion}

We need to obtain $\chi_\beta$, defined by
\begin{equation}
    \bm\nabla^2 \chi_\beta(\bm r) = \psi_\beta(\bm r).
    \label{eq:chi_def_app}
\end{equation}
To invert the Laplacian, it is convenient to decompose the problem into angular scalar and vectorial sectors. We work in the frame centered on the droplet by setting $\bm X=\bm 0$ so that $r\equiv|\bm r|$.

\paragraph{Scalar sector.}
For scalar radial functions such as $\psi_R(\bm r)=\psi_R(r)$ and $\chi_R$, we follow Ref.~\onlinecite{cates2023}. The Laplacian for a scalar radial function $\chi_R(r)$ is
\begin{equation}
    \bm\nabla^2 \chi_R(r)= \chi_R''(r) + \frac{d-1}{r}\chi_R'(r).
    \label{eq:laplacian_radial}
\end{equation}
Multiplying $\psi_R=\bm \nabla^2\chi_R$ by $r^{d-1}$ gives
\begin{equation}
    \begin{split}
    r^{d-1}\psi_R(r)
    &= r^{d-1}\chi_R''(r)+ (d-1) r^{d-2}\chi_R'(r)\\
    &= \big(r^{d-1}\chi_R'(r)\big)'.
    \end{split}
    \label{eq:radial_poisson_step}
\end{equation}
Integrating twice yields the inverse-Laplacian representation
\begin{equation}
    \chi_R(r)=
    -\int_r^{r_{\max}} dr_2\int_0^{r_2} dr_1\left(\frac{r_1}{r_2}\right)^{d-1}\psi_R(r_1).
    \label{eq:inv_laplacian_radial}
\end{equation}
In $d>2$ we may safely take $r_{\max}\to\infty$. In $d=2$, a large-scale cutoff $r_{\max}$ must be retained, as discussed below.

Using $\psi_R(r)=-g'(r-R)=\Delta\rho \delta(r-R)$,
Eq.~\eqref{eq:inv_laplacian_radial} gives
\begin{equation}
\begin{split}
\chi_R(r)&=
\Delta\rho\int_r^{r_{\max}} \int_0^{r_2} \left(\frac{r_1}{r_2}\right)^{d-1}\delta(r_1-R)dr_1dr_2\\
&=
\begin{cases}
-\dfrac{\Delta\rho R^{d-1}}{(d-2) \max(r, R)^{d-2}}, &  d>2,\\
-\Delta\rho R\log \left(\dfrac{r_{\max}}{\max(r, R)}\right), & d=2.
\end{cases}
\end{split}
\label{eq:chi_R_app}
\end{equation}
In particular, $\chi_R(R)=-\Delta\rho R/(d-2)$ for $d>2$ and $\chi_R(R)=-\Delta\rho R\log(r_{\max}/R)$ for $d=2$.

\paragraph{Vector sector.}
The translation mode is vectorial:
\begin{equation}
    \psi_{X_i}(\bm r)=\partial_{X_i}\rho(\bm r)=-g'(r-R) n_i,
    \qquad n_i\equiv {r_i}/{r}.
\end{equation}
Thus Eq.~\eqref{eq:inv_laplacian_radial} does not apply. Rotational symmetry implies that the corresponding solution must have the same angular dependence, so we set
\begin{equation}
    \chi_{X_i}(\bm r)=n_ih(r).
    \label{eq:chi_app}
\end{equation}
Using $\bm\nabla^2 n_i = -{(d-1)}n_i/{r^2}$, we obtain
\begin{equation}
    \bm \nabla^2 \big(n_i h(r)\big)=
    n_i\left(h''(r)+\frac{d-1}{r}h'(r)-\frac{d-1}{r^2}h(r)\right).
\end{equation}
Therefore $\bm \nabla^2\chi_{X_i}=\psi_{X_i}$ reduces to the radial ODE
\begin{equation}
    h''(r)+\frac{d-1}{r}h'(r)-\frac{d-1}{r^2}h(r)=-g'(r-R).
    \label{eq:h_ode}
\end{equation}
In the sharp-interface limit $g'(r-R)=-\Delta\rho \delta(r-R)$, solving Eq.~\eqref{eq:h_ode} using the same method as in Eq.~\eqref{eq:chi_R_app} with regularity at $r=0$ and decay at large $r$ gives
\begin{equation}
    h(r)= -\frac{\Delta\rho}{d}\left[r\Theta(R-r)+r(R/r)^d \Theta(r-R)\right],
    \label{eq:chi_app_h_value}
\end{equation}
and especially $h(R)=-\Delta\rho R/{d}$. 

\subsection{The matrix \texorpdfstring{$G_{\alpha\beta}$}{G}}

The values of $G_{\alpha\beta}$ are now straightforward to obtain using Eq.~\eqref{eq:chi_R_app}:
\begin{equation}
    \begin{split}
    G_{RR}(R)&=\langle \chi_R,\partial_R\rho\rangle\\
    &=
    \begin{cases}
    -\Delta\rho^2 S_d \dfrac{R^{d}}{d-2},& d>2,\\
    -\Delta\rho^2 S_d R^{2}\log\left(\dfrac{r_{\max}}{R}\right),&d=2,
    \end{cases}
    \end{split}
\end{equation}
as well as Eqs.~\eqref{eq:chi_app} and \eqref{eq:chi_app_h_value} yielding:
\begin{equation}
    G_{XX}(R)=\langle \chi_{X_i},\partial_{X_i}\rho\rangle
    =-\Delta\rho^2 S_d \frac{R^{d}}{d^{2}}.
\end{equation}
with $S_d$ is the surface area of the unit sphere in $\mathbb R^d$ which arises from $\int d\Omega = S_d$ with $d\bm r = r^{d-1}drd\Omega$. Notably, $S_2=2\pi$ and $S_3=4\pi$.

\subsection{The deterministic term \texorpdfstring{$\langle \psi_\beta,\tilde\mu[\rho]\rangle$}{}}
\label{sec:deterministic_projection}

The deterministic term is $\langle \psi_\beta,\tilde\mu[\rho]\rangle$ with $\tilde\mu=\mu-\mu_s$ and $\mu_s=f'(\rho_s)$. Using $\psi_R=\partial_R\rho=-g'(r-R)$ and $\mu=f'(\rho)-\kappa \bm\nabla^2\rho$, we write
\begin{equation}
\begin{split}
\langle \psi_R,\tilde\mu\rangle
=&-\int g'(r-R)\Big[f'(g(r-R))-f'(\rho_s)\\
&-\kappa \bm\nabla^2 g(r-R)\Big]d\bm r\\
=&-\int g'(u)\Big[f'(g(u))-f'(\rho_s)-\kappa g''(u)\\
&-\kappa \frac{d-1}{r}g'(u)\Big]d\bm r,\\
=&-S_d R^{d-1}\int g'[f'(g)-f'(\rho_s)]du\\
&+ \kappa S_d (d-1)R^{d-2}\int g'^2du,
\end{split}
\end{equation}
with $u=r-R$ and where the term $ g' g''$ vanishes by integration by parts.
Evaluating the remaining integral gives
\begin{equation}
    \begin{split}
    \langle \psi_R,\tilde\mu\rangle=&-S_d R^{d-1}\Big[f(\rho_s)-f(\rho_n)+\Delta\rho f'(\rho_s)\Big]\\
    &+ S_d(d-1)\gamma_0 R^{d-2},
    \end{split}
\end{equation}
with 
\begin{equation}
\gamma_0=\kappa\int g'(u)^2du
    \label{eq:surface_tension_mean_field}
\end{equation} the \emph{mean field} surface tension~\cite{onuki2002phase}.

By isotropy, $\langle \psi_{X_i}, \tilde\mu\rangle=0$ since the nucleus is not driven.

\subsection{The Laplacian noise \texorpdfstring{$\langle \psi_\beta,\zeta\rangle$}{}}
The Laplacian noise induces the following noise $\langle \psi_\beta,\zeta\rangle$, for the dynamics of the collective variables. We recall that:
\begin{equation}
    \overline{\zeta(\bm r, t)\zeta(\bm r', t')}=\delta(t-t')\delta(\bm r-\bm r').
\end{equation}
Symbolic manipulations lead to:
\begin{equation}
    \overline{\langle \psi_\alpha,\zeta\rangle(t)\langle \psi_\beta,\zeta\rangle(t')}=\langle \psi_\alpha, \psi_\beta\rangle\delta(t-t').
\end{equation}
Using the mean field surface tension Eq.~\eqref{eq:surface_tension_mean_field} yields directly:
\begin{equation}
    \begin{split}
    \langle \psi_R,\psi_R\rangle=\langle \partial_R\rho,\partial_R\rho\rangle &= S_d R^{d-1}\frac{\gamma_0}{\kappa},\\
    \langle \psi_{X_i},\psi_{X_j}\rangle=\langle \partial_{X_i}\rho,\partial_{X_j}\rho\rangle &= \delta_{ij}\frac{S_d}{d}R^{d-1}\frac{\gamma_0}{\kappa}.
    \end{split}
\end{equation}
The cross-correlations are zero.

\subsection{The divergence noise \texorpdfstring{$\langle \bm\nabla\chi_\beta,\bm\xi\rangle$}{}}

The divergence noise $\langle \bm\nabla\chi_\beta,\bm\xi\rangle$ has correlations:
\begin{equation}
    \overline{\langle \bm\nabla\chi_\alpha,\bm\xi\rangle(t)\langle \bm\nabla\chi_\beta,\bm\xi\rangle(t')}=\langle \bm\nabla\chi_\alpha,  \bm\nabla\chi_\beta\rangle\delta(t-t').
\end{equation}
Integration by parts yields:
\begin{equation}
    \langle \bm\nabla\chi_\alpha,\bm\nabla\chi_\beta\rangle=-\langle \chi_\alpha,\bm\nabla^2\chi_\beta\rangle=-\langle \chi_\alpha,\psi_\beta\rangle=-G_{\alpha\beta}.
\end{equation}

\subsection{The final equations}
\begin{widetext}
\label{sec:final_eq_app_nop_capillary}
Putting everything together, we obtain:
\begin{equation}
\dot R=
\begin{cases}
\begin{aligned}[t]
&\dfrac{\gamma_0(d-1)(d-2)}{\Delta\rho^2 R}\left(\dfrac{1}{R_c}-\dfrac{1}{R}\right)+\sqrt{2 D_R(R)}  \eta_R(t)+\sqrt{2 D_R^{(b)}(R)}  \eta_R^{(b)}(t)
\end{aligned}
& (d>2),\\
\begin{aligned}[t]
&\dfrac{\gamma_0}{\Delta\rho^2 R \log(r_{\max}/R)}\left(\dfrac{1}{R_c}-\dfrac{1}{R}\right)+\sqrt{2 D_R(R)}  \eta_R(t)+\sqrt{2 D_R^{(b)}(R)}  \eta_R^{(b)}(t)
\end{aligned}
& (d=2),
\end{cases}
\end{equation}
with:
\begin{equation}
D_R(R)=
\begin{cases}
D \dfrac{\gamma_0}{\kappa S_d} \dfrac{(d-2)^2}{\Delta\rho^4} \dfrac{1}{R^{d+1}},\\
D \dfrac{\gamma_0}{\kappa S_2} \dfrac{1}{\Delta\rho^4}
\dfrac{1}{R^{3}\log^2(r_{\max}/R)},
\end{cases} \qquad
D_R^{(b)}(R)=
\begin{cases}
D^{(b)} \dfrac{d-2}{\Delta\rho^2 S_d} \dfrac{1}{R^{d}},
& \qquad(d>2),\\
D^{(b)} \dfrac{1}{\Delta\rho^2 S_2} \dfrac{1}{R^{2}\log(r_{\max}/R)},
& \qquad(d=2).
\end{cases}
\end{equation}
and
\begin{equation}
R_c=
\frac{\gamma_0(d-1)}{(\rho_n-\rho_s)f'(\rho_s)-\big(f(\rho_n)-f(\rho_s)\big)}.
\end{equation}
$\eta_R$ and $\eta_R^{(b)}$ are uncorrelated Gaussian noises with unit variance. For the translational mode, we likewise find:
\begin{equation}
\dot {\bm X}=\sqrt{2D_X(R)}\bm\eta_X(t)+\sqrt{2D_X^{(b)}(R)}\bm\eta_X^{(b)}(t),\quad
D_X(R)=D \frac{\gamma_0}{\kappa\Delta\rho^4S_d} \frac{d^3}{R^{d+1}},\quad
D_X^{(b)}(R)= D^{(b)} \frac{1}{\Delta\rho^{2} S_d} \frac{d^{2}}{R^{d}}.
\end{equation}
All noises are uncorrelated and centered Gaussian with unit variance.

\end{widetext}
\section{Integrals for capillary waves}
\label{app:integrals_capillary}
\subsection{\texorpdfstring{$\chi_{a_{\ell J}}$}{chi} and Laplacian inversion}

In general, the capillary modes are coupled, which makes the equations complicated. Therefore, we work in the linear regime by neglecting the term $\delta R$ as standard in capillary wave studies~\cite{Maire2025b}:
\begin{equation}
    \begin{split}
    -g'(u)/\Delta\rho&=\delta(u)=\delta(r-(R+\delta R))\\
    &= \delta(r-R) - \delta R \delta'(r-R) + \mathcal{O}(\delta R^2)\\
    &\simeq \delta(r-R)
    \end{split}
\end{equation}
We then obtain:
\begin{equation}
    \psi_{a_{\ell J}}=-g'(u)Y_{\ell J}(\bm \Omega)\simeq \Delta\rho\,\delta(r-R)Y_{\ell J}(\bm\Omega).
\end{equation}
and $\chi$:
\begin{equation}
    \bm\nabla^2\chi_{a_{\ell J}}=\psi_{a_{\ell J}},
\end{equation}
Using the properties of the hyperspherical harmonics~\cite{avery2017hyperspherical} and the same method as in Eq.~\eqref{eq:radial_poisson_step}, we find:
\begin{equation}
    \chi_{a_{\ell J}}(r,\bm\Omega)=
    -\frac{\Delta\rho R}{2\ell+d-2}
    \begin{cases}
    \left(\dfrac{r}{R}\right)^\ell Y_{\ell J}(\bm\Omega), &r\leq R,\\
    \left(\dfrac{R}{r}\right)^{\ell+d-2}Y_{\ell J}(\bm\Omega), & r>R,
    \end{cases}
\end{equation}
and especially:
\begin{equation}
    \chi_{a_{\ell J}}(R, \bm \Omega)= -\frac{\Delta\rho R}{2\ell+d-2} Y_{\ell J}(\bm\Omega)
\end{equation}

\subsection{The matrix \texorpdfstring{$G_{a_{\ell J}a_{\ell' J'}}$}{G}}

Due to the linearization around the spherical surface and the orthonormality of the hyperspherical harmonics, we must have
\begin{equation}
    G_{a_{\ell J}a_{\ell' J'}}=\delta_{\ell\ell'}\delta_{JJ'}G_{\ell \ell}(R),
\end{equation}
where we defined:
\begin{equation}
G_{\ell\ell}(R)\equiv\langle \chi_{a_{\ell J}},\psi_{a_{\ell J}}\rangle
= -\frac{(\Delta\rho)^2R^d}{2\ell+d-2}.
\end{equation}

\subsection{The deterministic term \texorpdfstring{$\langle \psi_{a_{\ell J}},\tilde\mu[\rho]\rangle$}{}}
\label{sec:det_capillary}

Instead of performing the full integral, since the deterministic term is equilibrium-like (variational), we can take a shortcut. For $\tilde\mu=\delta\tilde F/\delta\rho$, the chain rule gives the exact identity
$\langle\psi_{a_{\ell J}},\tilde\mu\rangle=\partial_{a_{\ell J}}\tilde F$. Moreover, the shape dependence of $\tilde F$ comes from the interfacial term $\gamma_0 A$ with $\gamma_0=\kappa\int g'(u)^2 du$. Expanding the area to quadratic order in $a_{\ell J}$ yields~\cite{lenz2003hexatic}
\begin{equation}
\begin{split}
\tilde F(R,\{a_{\ell J}\})\!=&\tilde F_0(R)+\frac{\gamma_0}{2}R^{d-3}\sum_{\ell\ge2}\sum_J\left[\lambda_\ell-(d-1)\right]a_{\ell J}^2\\&+\mathcal O(a_{\ell J}^3),
\label{eq:F_tilde_subtle}
\end{split}
\end{equation}
with $\lambda_\ell=\ell(\ell+d-2)$. Hence,
\begin{equation}
\langle\psi_{a_{\ell J}},\tilde\mu\rangle=\partial_{a_{\ell J}}\tilde F=\gamma_0R^{d-3}(\ell-1)(\ell+d-1)a_{\ell J}.
\end{equation}

Particular care is required here because Eq.~\eqref{eq:F_tilde_subtle} depends explicitly on $R$. Consequently, capillary-wave modes contribute to $\dot R$ through the relation $\langle \psi_R,\tilde\mu\rangle=\partial_R \tilde F$:
\begin{equation}
\langle \psi_R,\tilde\mu\rangle
\!=\!\partial_R \tilde F_0(R)
+ \frac{\gamma_0}{2}(d-3)R^{d-4}\sum_{\ell\ge 2,J} c_\ell a_{\ell J}^2
+ \mathcal{O}(\{a_{\ell J}^3\}),
\end{equation}
where $c_\ell=\lambda_\ell-(d-1)$. The first term reproduces the result obtained in Sec.~\ref{sec:deterministic_projection}, whereas the second term is new. Although the latter is of order $a_{\ell J}^2$ and therefore appears negligible within our approximation, it must be retained. If omitted, the capillary waves would be influenced by the evolution of $R$, while $R$ would not receive the corresponding feedback from $a_{\ell J}$, rendering the perturbation theory nonreciprocal and non-potential.

\subsection{The Laplacian noise \texorpdfstring{$\langle \psi_{a_{\ell J}},\zeta\rangle$}{}}

We use again $\overline{\zeta(\bm r,t)\zeta(\bm r',t')}=\delta(t-t')\delta(\bm r-\bm r')$ which implies:
\begin{equation}
\overline{\langle\psi_{a_{\ell J}},\zeta\rangle(t)\langle\psi_{a_{\ell'J'}},\zeta\rangle(t')}=\langle\psi_{a_{\ell J}},\psi_{a_{\ell'J'}}\rangle\delta(t-t').
\end{equation}
Using $\psi_{a_{\ell J}}=-g'(u)Y_{\ell J}(\bm\Omega)$ we obtain
\begin{equation}
\langle\psi_{a_{\ell J}},\psi_{a_{\ell'J'}}\rangle
=\delta_{\ell\ell'}\delta_{JJ'}\frac{\gamma_0}{\kappa}R^{d-1}.
\end{equation}

\subsection{The divergence noise \texorpdfstring{$\langle \bm\nabla\chi_{a_{\ell J}},\bm\xi\rangle$}{}}
With $\overline{\xi_i(\bm r,t)\xi_j(\bm r',t')}=\delta_{ij}\delta(t-t')\delta(\bm r-\bm r')$,
\begin{equation}
\overline{\langle\bm\nabla\chi_{a_{\ell J}},\bm\xi\rangle(t)\langle\bm\nabla\chi_{a_{\ell'J'}},\bm\xi\rangle(t')}
=\langle\bm\nabla\chi_{a_{\ell J}},\bm\nabla\chi_{a_{\ell'J'}}\rangle\delta(t-t').
\end{equation}
Integration by parts yields again the identity
\begin{equation}
\langle\bm\nabla\chi_{a_{\ell J}},\bm\nabla\chi_{a_{\ell'J'}}\rangle=-G_{a_{\ell J},a_{\ell'J'}}.
\end{equation}

\subsection{Final equation for capillary modes}

\begin{widetext}
The matrix $G$ is still diagonal due to our linearization, hence we can use Eq.~\eqref{eq:dynamics_slow_field_before_integral}. The modes evolve as:
\begin{equation}
\dot a_{\ell J}=-\Gamma_\ell(R)a_{\ell J}+\sqrt{2D_{\ell}(R)}\eta_{a_{\ell J}}(t)+\sqrt{2D_{\ell}^{(b)}(R)}\eta_{a_{\ell J}}^{(b)}(t),
\qquad (\ell\ge 2),
\end{equation}
with
\begin{equation}
\Gamma_\ell(R)=\frac{\gamma_0}{\Delta\rho^{2}R^{3}} (2\ell+d-2) (\ell-1)(\ell+d-1),
\end{equation}
and the effective noise strengths
\begin{equation}
D_{\ell}(R)=D \frac{\gamma_0}{\kappa} \frac{(2\ell+d-2)^2}{\Delta\rho^{4}}\frac{1}{R^{d+1}},\qquad
D_{\ell}^{(b)}(R)=D^{(b)} \frac{2\ell+d-2}{\Delta\rho^{2}} \frac{1}{R^{d}}.
\end{equation}
Here $\eta_{a_{\ell J}}$ and $\eta_{a_{\ell J}}^{(b)}$ are independent centered Gaussian white noises of unit
variance, and vanishing cross-correlations between the two noise families. We note that we restricted $\ell\geq2$ because $\ell=1$ is the translation mode and $\ell=0$ is the radial mode, already taken into account by the equation for $\dot R$.

We recall that in $d\neq3$, $\dot R$ is now affected by capillary waves via the feedback term found in Sec.~\ref{sec:det_capillary}.\\

\end{widetext}
\section{Mathematical peculiarities in \texorpdfstring{$d=2$}{d=2}}
\label{sec:d=2}

In $d=2$, the radial dynamics can also be written in the form
\begin{equation}
    \dot R=-\mathcal M_R(R)\dfrac{\partial \tilde W}{\partial R}+\sqrt{2D\mathcal M_R(R)}\eta_R(t),
\end{equation}
where $\tilde W$ and $\mathcal M_R$ are obtained directly from the expressions in Sec.~\ref{sec:final_eq_app_nop_capillary}:
\begin{equation}
    \begin{split}
        \mathcal M_R&=\dfrac{\gamma_0}{2\pi\Delta\rho^4\kappa R^3\log\left(\dfrac{r_{\max}}{R}\right)^2},\\
        \tilde W(R)&=\frac{\kappa \Delta\rho^2\pi}{18} R^3\!\left[\!\!\dfrac{9}{R} \!-\! \dfrac{4}{R_c}\! +\! 6\left(\dfrac{3}{R} - \dfrac{2}{R_c}\right)\!\log\left(\dfrac{r_{\max}}{R}\right)\!\right]\!.
    \end{split}
\end{equation}

As in $d=3$, $\tilde W$ contains the expected $R^d$ and $R^{d+1}$ contributions, but in $d=2$ it also acquires a logarithmic correction that depends on $r_{\max}$. Evaluated at the critical radius, the barrier
\begin{equation}
\tilde W(R_c)=\frac{\Delta\rho^2\kappa\pi}{18}R_c^2\left[5+6\log(r_{\max}/R_c)\right]
\end{equation}
grows with $r_{\max}$, which naively suggests a suppression of nucleation in larger systems. At face value, this is reminiscent of the suppression reported in Ref.~\onlinecite{Lei2023}. However, that work concerns nucleation into an absorbing phase, which is qualitatively different from the liquid--gas nucleation problem considered here.

Moreover, because the divergence is only logarithmic, the associated suppression is only algebraic in system size. Indeed, near $R\simeq R_c$, the dominant factor behaves as
\begin{equation}
e^{-\tilde W(R_c)/D}\sim \left(\frac{r_{\max}}{R_c}\right)^{-\alpha},
\end{equation}
for some $\alpha>0$, rather than exponentially in $r_{\max}$. In equilibrium systems with short-range correlations, such an algebraic suppression can be offset by the growth in the number of statistically independent nucleation regions, which scales as $(r_{\max}/\xi)^d$ with correlation length $\xi$. In our case, however, this counting argument is not straightforward, because the hyperuniform state is long-range correlated and independent nucleation volumes are not clearly defined.

A second reason for caution is that the $r_{\max}$ dependence may be an artifact of the single-droplet description. In an infinite system, one expects multiple nuclei to grow simultaneously; in that regime, $r_{\max}$ should be interpreted not as the system size, but as an effective screening length set by the typical interdroplet separation~\cite{Marqusee1984,zheng1989theory}. 

Finally, within the Stratonovich discretization, the stationary distribution also contains a subexponential prefactor (even at equilibrium),
\begin{equation}
    P(R)\sim \dfrac{1}{\sqrt{\mathcal M_R(R)}}e^{-\tilde W(R)/D},
\end{equation}
with $1/\sqrt{\mathcal M_R(R)}\sim \log(r_{\max}/R)$. This prefactor, however, increases with $r_{\max}$, and only logarithmically. It is therefore subdominant compared to the algebraic suppression coming from $\tilde W$. 

Taken together, these observations suggest that the explicit $r_{\max}$-dependent corrections are most likely artifacts of the present coarse-grained theory. Especially since preliminary molecular dynamics simulations show no clear dependence of the observed distribution of clusters on system size.

\section{Relaxation and activated dynamics of \texorpdfstring{$R$ and $a_{\ell J}$}{R and alJ}}
\label{app:relaxation}

In this appendix, we show that, under the physically proper definition of a nucleation event, the capillary-wave dynamics along the instanton are purely relaxational in $d=3$. We define a nucleation trajectory $\bm q^{\rm nuc}(t)$ as \emph{any} trajectory starting at \emph{any} point in the metastable basin $\bm q(0)\in \mathcal B$ and reaching $R(t_c)=R_c$, with unfixed $\{a_{\ell J}\}$. This should not be confused with $\bm q^{\rm nuc}_*(t)$ used in the main text, where the initial point is fixed exactly, and the final capillary-wave amplitudes are also prescribed.

In the hyperuniform case ($D^{(b)}=0$), Eq.~\eqref{eq:coupled_neq} can be written as
\begin{equation}
\begin{split}
    \dot q_\beta&=b_\beta(\bm q)+\sqrt{2D\mathcal M_\beta(\bm q)}\eta_\beta(t),\\
    b_\beta(\bm q)&\equiv -\mathcal M_\beta(\bm q)\partial_\beta \tilde U_\beta(\bm q).
\label{eq:langevin_b}    
\end{split}
\end{equation}
The most probable path, in the weak-noise limit, is the minimizer of the Freidlin-Wentzell action $\mathcal A$~\cite{Freidlin1984}
\begin{equation}
    \dfrac{\mathcal A}{D}=\int_0^{t_c}L(\bm q,\dot{\bm q})dt,\quad L(\bm q,\dot{\bm q})=\frac{1}{4}\sum_\beta\frac{\big(\dot q_\beta-b_\beta(\bm q)\big)^2}{\mathcal M_\beta(\bm q)},
    \label{eq:FW}
\end{equation}
under the boundary conditions:
\begin{equation}
\bm q(0)\in \mathcal B,\qquad R(t_c)=R_c.
\end{equation}

It is convenient to derive the optimal-path equations in Hamiltonian form~\cite{avanzini2024methods}:
\begin{equation}
H(\bm q,\bm p)=\sum_\beta p_\beta \dot q_\beta -L=\sum_\beta\left[\mathcal M_\beta(\bm q)p_\beta^2+b_\beta(\bm q)p_\beta\right],
\label{eq:hamiltonian_path}
\end{equation}
with the conjugate momenta
\begin{equation}
p_\beta\equiv \frac{\partial L}{\partial \dot q_\beta}=\frac{\dot q_\beta-b_\beta}{2\mathcal M_\beta}.
\label{eq:momentum}
\end{equation}
The minimizing paths obey Hamilton's equations
\begin{equation}
\begin{split}
\dot q_\beta&=b_\beta(\bm q)+2\mathcal M_\beta(\bm q)p_\beta,\\
\dot p_\beta&=-\partial_\beta H.
\end{split}
\label{eq:hamilton_eq}
\end{equation}

The conjugate momentum $p_\beta$ measures how much the optimal path departs from the deterministic drift, since
\begin{equation}
p_\beta=\frac{\dot q_\beta-b_\beta}{2\mathcal M_\beta}.
\end{equation}
Thus $p_\beta=0$ corresponds to the purely deterministic branch $\dot q_\beta=b_\beta$, while $p_\beta\neq 0$ corresponds to an activated fluctuation.

To cross the barrier, $R$ must reach $R_c$, which requires a nonzero momentum along the optimal path. By contrast, for variables whose final value is not constrained, such as capillary modes $a_{\ell J}$, we have to set
\begin{equation}
p_{a_{\ell J}}(t_c)=0.
\label{eq:p_a=0}
\end{equation}
to satisfy the Euler-Lagrange equation,
\begin{equation}
    \delta \mathcal A=\Big[\sum_\beta p_\beta\,\delta q_\beta\Big]_{0}^{t_c}+\int_0^{t_c}\sum_\beta\left(\frac{d}{dt}\frac{\partial L}{\partial \dot q_\beta}-\frac{\partial L}{\partial q_\beta}\right)\delta q_\beta\,dt,
\end{equation}
since $\delta q_{a_{\ell J}}(t_c)\neq 0$. Eqs.~\eqref{eq:p_a=0} and \eqref{eq:hamilton_eq} imply
\begin{equation}
    p_{a_{\ell J}}(t)=0,\quad t\in[0, t_c],
\end{equation}
hence, during nucleation, the capillary modes \emph{relax} deterministically while $R(t)$ is activated. Indeed, using the Hamilton-Jacobi equation~\cite{avanzini2024methods} $H(\bm q, \bm p)=0\Rightarrow p_R = -b_R/\mathcal M_R$, we find:
\begin{equation}
    \dot R^{\rm nuc}(t) = -b_R(R^{\rm nuc}),\quad \dot a_{\ell J}^{\rm nuc}(t) =b_{a_{\ell J}}(R^{\rm nuc}, a_{\ell J}^{\rm nuc}),
    \label{eq:nuc}
\end{equation}
while the relaxational dynamics are:
\begin{equation}
    \dot R^{\rm rel}(t)  = b_R(R^{\rm rel}),\quad \dot a_{\ell J}^{\rm rel}(t) =b_{a_{\ell J}}(R^{\rm rel}, a_{\ell J}^{\rm rel}).
    \label{eq:rel}
\end{equation}

Therefore, the capillary modes $a_{\ell J}$ play no essential role; in both cases, they simply relax to zero, and the upward path of $R$ is given by $-b_R$, as in equilibrium. The nucleation problem is thus more transparently described in terms of the single coordinate $R$, whose optimal path is the time reverse of the relaxational trajectory.

We emphasize, moreover, that within this class of trajectories it is not meaningful to search for violations of detailed balance, since the paths are not defined by fixed endpoint constraints.

\end{document}